# Advanced Personnel Vetting Techniques in Critical Multi-Tennant Hosted Computing Environments

Farhan Hyder Sahito

Institute for Software Technology
Graz University of Technology
Graz, Austria

Wolfgang Slany

Institute for Software Technology
Graz University of Technology
Graz, Austria

*Abstract*—**The emergence of cloud computing presents a strategic direction for critical infrastructures and promises to have far-reaching effects on their systems and networks to deliver better outcomes to the nations at a lower cost. However, when considering cloud computing, government entities must address a host of security issues (such as malicious insiders) beyond those of service cost and flexibility. The scope and objective of this paper is to analyze, evaluate and investigate the insider threat in cloud security in sensitive infrastructures as well as to propose two proactive socio-technical solutions for securing commercial and governmental cloud infrastructures. Firstly, it proposes actionable framework, techniques and practices in order to ensure that such disruptions through human threats are infrequent, of minimal duration, manageable, and cause the least damage possible. Secondly, it aims for extreme security measures to analyze and evaluate human threats related assessment methods for employee screening in certain high-risk situations using cognitive analysis technology, in particular functional Magnetic Resonance Imaging (fMRI). The significance of this research is also to counter human rights and ethical dilemmas by presenting a set of ethical and professional guidelines. The main objective of this work is to analyze related risks, identify countermeasures and present recommendations to develop a security awareness culture that will allow cloud providers to utilize effectively the benefits of this advanced techniques without sacrificing system security.**

## I. INTRODUCTION

Cloud computing is the crest of the wave of new and better computing possibilities – provides, efficient and cost effective way to run governmental and commercial organizations. This technology is unique in its ability to address our national defense [1]. It offers critical infrastructures and intelligence agencies an emerging platform to deliver innovative mission solutions and assist in enhanced collaboration [2]. This platform underpins the next generation of digital products and services with reducing time and decreasing cost and giving these agencies the ability to purchase a broad range of IT services in a utility- based model to increase their operational efficiencies [3]. However, despite the myriad advantages of cloud computing, several security challenges still exist, such as malicious insider [4]. Insiders, by virtue of legitimate access to their organizations' systems and networks, may pose a significant risk to critical infrastructure in cloud. It poses a significant risk and could gain complete control over the cloud services with little or no risk of detection. This paper will focus on malicious insider as a major threat in cloud security.

Recent reports[1] expose that violent extremists are trying to obtain insider positions in critical infrastructures. Based on these reports, it is clear malicious insiders pose a significant threat to cloud computing and critical agencies and government organizations are careful in adopting this new IT approach as moving to cloud could be challenging with due consideration to the sensitivity of data [1]. Cloud providers have an extreme interest in detecting malicious insiders to assure the protection of critical infrastructures. Despite much investigation into the motivation and psychology of malicious insiders, the fact remains that it is extremely complicated to predict this threat [5]. This presents cloud vendors with a dilemma to establish an appropriate level of trust w.r.t. employees. The objective of this work is twofold and the purpose is to alleviate above threats by focusing on two proactive approaches to overcome the aforementioned issues.

The goal of the first section is to develop a framework that focuses on a multi-layered security strategy that can be used to better combat the risks of the insider threat to an organization's networks, or information systems. This paper recommends that this strategy might be a necessary tool for cloud providers, policy makers, security officers and other stakeholders to identification and assessment of risk that insiders presents to organizations. Innovations in techniques and framework will enable cloud providers to successfully execute the mission of cloud computing. The next session is aiming for extreme measures to evaluate human threats related assessment methods for employee screening and evaluations using cognitive analysis technology, in particular functional Magnetic Resonance Imaging (fMRI). The aim is to establish an appropriate level of trust at employees, effective monitoring and ensuring that insiders do not pose a foreseeable risk to critical infrastructure. The technological and non-technological outputs will ensure a better involvement of SME's as well as enhancing the competitiveness that enables users, ranging from SME's to government agencies, to experience a secure and trustworthy cloud computing experience.

## II. MALICIOUS INSIDERS IN CRITICAL INFRASTRUCTURE - A THREAT TO CLOUD SECURITY

Critical infrastructures are the advanced physical and cyber-based systems essential to the state's security, economic prosperity and social well-being of the nation, such as law enforcement services, power plants and Information and communication services etc. [6]. As a result of advances in technology, these critical infrastructures have become increasingly automated and interlinked. On the other side, these advances have created new vulnerabilities to physical and cyber-attacks by insiders. A report [2] issued by the US Department of Homeland Security revealed that violent extremists are trying to obtain insider positions in critical infrastructure in the USA and the world and may increase the impact of any attack on the government assets. It further reveals that the fall edition of AQAP (a magazine published by Al-Qaeda) encourages followers to use "specialized expertise and those who work in sensitive locations that would offer them unique opportunities" to conduct terrorist attacks in the world.

---

[1] http://info.publicintelligence.net/DHS-InsiderThreat.pdf

A recent study[3] "2011 Cost of Data Breach Study: United States" reveals that insiders are the top cause of data breaches and 25 percent more costly than other types. Current events also expose that concerning cloud computing in the context of critical infrastructure, the threat of malicious insiders is very real. The cloud computing architecture is primarily a multi-tenant, service based architecture that necessitates the creation of certain staff positions that can be a high risk in terms of internal security threats [2]. Moving data and application in cloud environment bring with it an inherent level of risk that allows insiders to steal the confidential data (such as passwords, cryptographic keys, clear text passwords from a VM's memory, private keys from a VM's memory etc.) of cloud users, sabotage of information resources and various types of frauds [4]. A cloud administrator's access to the management VM makes these attacks possible [4]. Recently, a denial of service attack launched by a malicious insider against a well know Infrastructure as a Service (IaaS) cloud by creating twenty accounts and launching virtual machine instances for each. Using those accounts, malicious insider created twenty additional accounts machine instances in an iterative fashion and thus consumed resources beyond set limits [7].

There have been several recent events that exhibit how human operators undermined cloud computing. In December 2010, Google has revealed that the accounts of Chinese human rights activists in Google's clouds were targeted in a hacking attempt. As part of this investigation it was discovered that these accounts were not compromised through a security breach at Google, but most likely by a malicious insider. In another case[4], David Barksdale, a 27-year-old former Google engineer, repeatedly took advantage of his position as a member of an elite technical group at the company to access users' accounts, violating the privacy of at least four minors during his employment. Security breaches by human elements in clouds are on the increase globally and there have been similar attacks at, e.g., RSA, Lush, Play.com and Epsilon around the world[5]. The sequence of scandals induced 2010 by the publication of classified government documents to the Wiki-Leaks website – in which volumes of sensitive documents were leaked by a trusted insider and ultimately published on an open website – has caused much embarrassment to the United States and other nations and represents the ultimate nightmare scenario for cloud vendors when considering the human aspect in cloud security [8]. It is indeed sobering to imagine that any organization could fall victim to such events and the damage cloud insider can do [8].

The exact security concerns rose from the LAPD (Los Angeles Police Department) when they decided to move to Google apps in November 2009 [7]. The security requirements included encryption of all data and background checks on employees who access the police database (e.g. criminal history records and fingerprints) could be vulnerable by malicious employees [7]. This data base is accessible to all police officials around the country. After more than two years of efforts, the city of LA has abounded plans in January 2012 and voted to scale back the contract to migrate its 13,000 law-enforcement employees to office application platform and Google's hosted email, saying that security needs of crucial departments were not able to be met and concluded that cloud provider could not be brought into compliance with certain security requirements of the FBI's Criminal Justice Information Systems[6]. Instead, Google will pay up to $350,000 per year for the LAPD for the entire term of the contract due to the delay.

Cloud providers need to face the reality that these threats are targeting critical infrastructure that could debilitating impact on state's defense and a loss of public confidence in state's services [3]. This could be a sign of growing pains as leading cloud providers are trying to get larger federal and government entities to go to the cloud and it also brings up questions about whether cloud service is really secure enough.

### III. MALICIOUS INSIDER IN CRITICAL INFRASTRUCTURE: WHY CANNOT WE STOP IT?

The "insider" is an individual authorized to access an organization's information system, network or data - based on trust [9]. The insider threat refers to harmful acts and malicious activities that trusted insiders might carry out such as negligent use of classified data, unauthorized access to sensitive information, fraud, illicit communications with unauthorized recipients, and any other behavior that cause harm to the organization [10]. Insiders can be system administrator, contractors, former employees, suppliers, security guards and partner employees etc. According to Noonan & Archuleta [11] malicious insiders can be labeled as three different types of actors: 1) criminals 2) ideological or religious radicals; and 3) psychologically-impaired disgruntled or alienated employees. The motivations of malicious insiders can be simple illicit financial gain, revenge for a perceived wrong, or a

---

radicalization for the advancement of religious or ideological objectives [11].

To counter human threats, agencies have invested billions of Euros in different technical measures for years now [12]. The current security paradigms include access control and encryption to face malicious insiders and outsiders. They are implemented through passwords, physical token authentication and biometric authentication, firewalls, encrypted data transmission, data leakage prevention, behavioral-pattern threat detection, voice stress analysis and polygraphs. Various studies demonstrate that above devices and security software are normally designed to defend against external threats to secure critical infrastructure and do not protect against attacks aided by internal help in organizations [12]. An insider not only has the ability to obtain or access valuable data that resides on the internal network, but he/she can obtain this data from their workstation without causing suspicion or breaking trust. Cloud organizations are well aware of these issues, as demonstrated in a recent roundtable including senior staff as well as the sector's major companies [13]. It is of utmost importance to adopt extreme measures to secure critical infrastructure in clouds to lower the level of threats. Secondly, well thought-out policies are necessary with understanding of the potential role of a process and methods – each designed to address threats targeted at specific segments of the cloud environment. Our research aims at alleviating this critical factor with higher assurance of achieving their desired security as employees at critical positions are may be, in some instances, the first and only line of defense and vital to national security.

## IV. MITIGATING INSIDER THREATS - TOWARD BEST PRACTICE ACTIONABLE FRAMEWORKS

This part focuses on malicious insiders with a variety of specialized security solutions in supporting all aspects of user trust in clouds to mitigate human threats and for making better informed decision when choosing a cloud solution. Our actionable framework aims to mitigate the potential vulnerabilities of critical infrastructure in cloud computing and key resources to ensure its protection and resilience. To achieve this goal, this paper recommends policies and actions to detect and manage insider threats and effectively fight these problems.

### A. A Human Factor Vulnerability Assessment (HVA) Framewor.

To mitigate malicious insider risks, this research recommends developing a human factor vulnerability assessment (HVA) framework. The purpose is to assess cloud organizational structure, its infrastructure, technical and non-technical vulnerabilities, employee roles and responsibilities, actual events of human threats incidents and existing security measures. This framework will raise awareness of the malicious insider causes, potential indicators and prevention and detection strategies, and informing cloud providers and users of their security responsibilities. This research recommends to build a data base in HVA that will consist of various incident for malicious related activates, vulnerabilities, lessons learned, and physiological profiles or statistics regarding the insider and insider misuse, or social engineering activities in cloud computing. This assessment may provide indicators, practices, actions, policies and procedures so cloud organization can track changes in their capabilities related to human threats over time and possible mitigation strategies. The assessment's results will benefit all individuals involved in this process and enable cloud providers to gain a better understanding of vulnerabilities in their clouds. The framework must include low-cost, easily implemented policy solutions for cloud vendors that have long term effects. These security awareness and training programs will be paramount to ensure that cloud providers and users will understand their security responsibilities to mitigate social engineering threats, and properly use and protect the cloud resources entrusted to them. This security plan ought not to be static; it has to be refined and adjusted regularly as the security challenges of cloud data centers permanently will change.

### B. Counter-Insider Threat Database

This database may include cloud users, security personnel, technical/non-technical staff, administrators, and all levels of organization management into a single, actionable assessment framework. Based on this database, HVA will analyze the newest approaches to topics related to human factors involved in cloud security by means of reviewing relevant publications and security surveys. It may also take into consideration the incidents knowledge and the opinion of security experts, psychologists and stakeholders, in order to find the real nature and magnitude of the malicious insider and social engineering problems to identify the best practices including individual and organizational actions and responses. It is important to compile a database in this phase with criminal cases in which current or former employees, contractors, or business partners abused the trust and access associated with their positions. This database would work on cases of insider activity from the real world documented with many insider threat cases that may provide a rich source for empirical research on real cases of insider threat. In this regard, interviews with various victim organizations as well as with perpetrator are necessary, complementing a wealth of case data with first-hand insights into the methods and motivations behind these crimes.

Secondly, the collaboration with noted psychologists and others from the law enforcement agencies to uncover key technical, social, and organizational patterns of insider behavior may severely hamper understanding of the magnitude of the problem and development of solution strategies. Insider misuse, abuse and malicious activates are yet another manifestation of betrayal of trust behavior for which cloud organization must be alert in espionage cases. The psychological profile will provide managers, security specialists and medical personnel a profile of the insider, which may become a useful tool to enable them to identify potential abusers before they cause serious damage This knowledge will investigate the objectives, data-driven examination of the motivations and behavior patterns of malicious/outsider human threats, as well as organizational issues that may influence them. Leveraging data to prevent, detect, and respond to these threats can help cloud organizations to strengthen the protection of the critical infrastructure in cloud. Potential benefits from developing an insider and outsider event database will focus on the need for improved detection, technical research priorities, and prevention through policies, education and training. This valuation may provide reports (confidential and public) on the basis of this assessment with the observable indicators and information and guidelines that cloud providers need to develop, and a plan of action and security standards to increase their ability to prevent, detect and respond to human threats. A unified database may be built that may consist of:

*a) Incidents*

*b) Statistical findings and implications regarding technical details of the incidents*

*c) Insider Planning*

*d) Nature of harm*

*e) Communication behavior and characteristics*

*f) Vulnerabilities*

*g) Lessons learned*

*h) Physiological profiles*

*i) Statistics regarding the insider and insider misuse or abuse*

*j) Case studies regarding detection and identification of the insiders*

*k) Social engineering activities to develop recommendations for technology and policy solutions for future problems in cloud computing environment.*

At minimum, the following activities should be engaged for a Human Factor Vulnerability Assessment (HVA) framework:

*1) Identify categories of problems and analyze differences/similarities of cases.*

*2) Provide administrators, security specialists a physiological profile of the insider/social engineer.*

*3) To define significant characteristic types of insider misuse and assist in the development of questions for security investigations.*

*4) Develop standardized framework and material for security education, awareness and training.*

*5) Provide data for finer-grain access policies and differential access controls needed to help define what constitutes proper usage, thus facilitating the role of insider-misuse detection.*

*6) Develop recommendations for technology and policy solutions for future problems.*

The next sections will define how this framework will deliver organizations the means to support business continuity by better securing their assets through well founded decision making and therefore lower the overall risks associated with cloud computing.

## V. Multi-Layered Security Strategy

This task may carry out a multi layered security strategy based on the Human Factor Vulnerability Assessment (HVA).

After this initial step, evaluation of the security mechanisms must be developed to monitor compliance and effectiveness of this task to revise the specification and architecture according to the experimental results obtained with the database. Our security program identifies the four critical steps:

1. Human Factor Security Awareness and Training Program Development

2. Security Awareness and Training Material Development

3. Security Program Implementation Consultancy

4. Post-Implementation Security Program Consultancy and Periodical Evaluation

*A. Human Factor Security Awareness and Training Program Development:*

This step must identify:

*a) What are the plans for developing and implementing security awareness and training opportunities to mitigate the human factors in cloud that are compliant with existing directives?*

*b) What awareness, training and education are needed to mitigate the human factors in clouds (i.e., what is required)?*

*c) How are these needs being addressed by cloud providers?*

*d) Where are the gaps between the needs?*

*e) What is being done and what more needs to be done for gap analysis and targeting deficient areas for early rollout?*

*f) Which needs are most critical in cloud environment?*

*g) What are the roles and responsibilities of a cloud organization's personnel will be identified in design and implementation stage to maintain security standards?*

*h) Who should ensure that the appropriate employee attend or view the applicable material in cloud agency?*

*i) Documentation, feedback, and evidence of learning for each aspect of the program;*

*j) Gap analysis and targeting deficient areas for early rollout.*

*B. Security Awareness and Training Material Development:*

In this task we recommend to develop a standardized framework and material to identify:

*a) What skills we do want the cloud provider's management and their employees to learn and apply to mitigate human threats?*

*b) What behavior to reinforce?*

*c) The topics may be covered in this approach are:*

*1) Social engineering tricks*

*2) Malicious insiders*

*3) Shoulder surfing*

*4) Dumpster diving*

*5) Tailgating*

*6) Access control issues*

*7) Visitor control*

*8) Physical access to spaces as well as handling incident response.*

*C. Security Program Implementation:*

This approach will focus on the implementation of a multi layered strategy with:

*1) Trainings*

*2) Programs*

3)  *Education and awareness.*

4)  *Trainings:*

This scope of this section is to ensure that cloud providers and users are appropriately trained in how to face human factors in cloud environment. It must ensure that insiders with privileged access in cloud such as administrator are only required to undergo the same vetting procedures as other insiders. It is evident that the sensitivity of functions they perform and the potential to access the most sensitive information contained in the system are much greater than those without such privileges. It is particularly important to focus on developing a strong security partnership with system managers, ensuring that these individuals receive the best security awareness training available. Furthermore, it is important to ensure that:

*a) Awareness and training material is effectively deployed to reach the intended audience.*

*b) Training material is reviewed periodically and updated when necessary and assist in establishing a tracking and reporting strategy.*

5)  *Programs:*

Security programs should recommend best methods, guidelines and technologies dealing with human factor issues. Some of our guidelines are:

*a) To recommend technologies and guidelines currently available for dealing with the insider/outsider problem.*

*b) To establish an activity to evaluate on a continuing basis the effectiveness of available security methods and tools of all types that may be available to mitigate that risk.*

*c) To direct the appropriate cloud providers to accelerate the development of new tools for cloud computing to combat insider and outsider threats.*

*d) To develop robust policies to discarded floppy disks, and printouts etc.*

*e) To propose different policies that classifies the sensitive information and enforces the mandatory and discretionary access control mechanisms.*

6)  *Education and Awareness*

This research identifies that many cloud operators lack an appropriate awareness of the threat insiders and social engineers pose to their operations. A strong security partnership is necessary to be developed. Education and awareness may present the biggest potential return for policy by motivating cloud operators and focusing their efforts to address the human threat. Our research attempts to enforce knee jerk policies to require all employees review the company policy after every three months, to acquaint themselves with revisions if any. Different activities are needed to be established to evaluate on a continuing basis the effectiveness of available security methods and tools of all types to mitigate that risk. Methods and techniques must be developed regarding discarded storage medium that may contain sensitive information. Appropriate awareness will help to shape the human threat policies and programs needed to address the unique insider risk profile.

***Techniques for Delivering Awareness Material:***

Techniques may include, but not limited to:

*a) Teleconferencing sessions*

*b) Interactive video training*

*c) Web based training*

*d) Videos, posters ("do and don't lists" or checklists)*

*e) Screensavers*

*f) Warning banners*

*g) SMS/messages*

*h) In-person*

*i) Instructor-led sessions*

*j) Awards program*

*k) Letters of appreciation etc.*

STAKEHOLDER WORKSHOP: It is recommended to arrange workshops for stakeholders involved in this whole process for intermediate results and panel discussion for feedback.

PUBLIC WORKSHOP: Arrange workshops to recommend security personnel and cloud users a layered defense through use of guidelines and tools and how to develop an effective, comprehensive insider/outsider threat monitoring strategy.

VI.  WHO WILL WATCH THE WATCH MAN?

In addition to above policies and framework, this paper suggests that it's a mistake to rely just on preventive measures. Instead, it is essential to supplement them with monitoring and auditing so insider attacks can be detected and truly stopped by removing the attacker from the cloud organization. Although it's difficult to prevent a malicious attack from a motivated insider, there are ways to spot bad behavior before it becomes a big problem. IT administrators make the cloud service and related datacenters operate, and through homogeneity and greater automation they often manage ratios of thousands of servers. The risk from these privileged cloud provider administrators must be explicitly recognized and addressed. There is a need to ensure that important location and data should not be reached and accessed by an individual and functions are not held by the same individual. Checks and balances implemented through review and approval processes are consistently applied so that gaps, even in critical situations are appropriately controlled and reviewed. Increased attention must be paid to privileged accounts, as it is mentioned earlier that how privileged administrators misused user's data and applications by accident, for profit, or for retribution. Each employee has logical patterns of information usage, and the organization should look for abnormal usage and investigate when this occurs. For example, if an employee looks at 50 customer accounts each day and then one day looks at 100 or more, there is a potential issue that should be investigated. You always need to understand if unusual behavior is warranted or malicious. Our strategy suggests that special monitoring and double auditing should be done to monitor abnormal or suspicious behavior by administrator or the damage that can be caused.

However, we advocate that in the context of critical infrastructure, employment screening measures are necessary to implement by focusing on technical and psychological measures used in psychological research as well as in law enforcement domain to secure cloud data. This research propose that technical measure like, fMRI scanning must be used for critical staff to detect malicious intent activities and susceptible behavior and other weaknesses (for instance, alcoholic, religious affiliation or domestic problems). This policy is needed to be implemented on pre-employment screening as well as regular employees. We propose that pre scanning measures such as reference checks can uncover prior criminal records, issues with character or credit problems.

## VII. MAINTAINING SECURITY USING FUNCTIONAL MAGNETIC RESONANCE IMAGING (FMRI)

The functional MRI is widely known and accepted in the scientific community as it does have a significant amount of scientific research behind its claims and validity. This technique relies on the fact that cerebral blood flow and neuronal activation are coupled. It involves placing the subject in a donut-shaped magnetic technology, which can identify subtle changes in electromagnetic fields [14]. During scanning, when an area of the brain is in use, blood flow to that region also increases [15]. Thus, its responses associated with neuronal activation correlate with cognitive tasks and various behavioral functions [15].

An fMRI has already had a major impact on neuroscience and in clinical settings. It has been applied ranging from language comprehension to treatment of neurological impairment disease, psychiatric illness, aesthetic judgment and justification of cognitive enhancing drugs in educational settings [16]. With these rapid developments many researchers claimed this technology to be useful outside the laboratory settings. For instance, economic contexts, investing personality traits, mental illness, religious extremism, racial prejudice, suicidal thoughts aggressive or violent tendencies and truth verification [16] [17]. Proponents of this neuro-imaging technology hailed fMRI as the next "truth meter" and conclude that because of the novelty of the physiological parameters being measured, this technology may be more accurate than other traditional methods for employee screening (e.g., polygraph, see [18] [19] [20] [21]). According to researchers, this ground breaking research proves that fMRI has the capacity to address the question of guilt versus innocence. Since the first publication by [21] on deception detection by fMRI, various papers and studies (See [22] [23] [24] [25] [26] [27] [28] [29] [30] [31] [32] [33] [33]) have reported different experiments in which subjects were asked to respond deceptively in some blocks and truthfully in others. It proved that lying involve more efforts than truth and expose that specific brain areas respond strongly in generating deceptive responses. As with lying, several brain regions show significant increases and light up on during scanning when a person sees a familiar object or image or during deception compared to truth telling [23]. For instance, dorsolateral prefrontal cortex (DLPFC), anterior cingulate cortex (ACC), ventrolateral (VLPFC) and left and right cerebral hemispheres increases activity when people tell lies [31].

FMRI truth verification in individual humans has been studied in 31 original peer reviewed scientific journal articles. These studies were done by 22 different research groups that included researchers from 13 different countries (USA, UK, Canada, Australia, China, Japan, Netherlands, Switzerland, Poland, Denmark, Sweden, Germany, and Russia). Similarly, during the employee screening phase, if any suspect employee is asked a question, the information to which is unknown then the specific regions of the brain is unusually active and it is presumed that suspect is lying; if, however, the same areas are no more active it may presumed that subject is telling the truth [18] [19] [20]. Thus, this technology has potential to reveal recognition regardless of whether the suspect speaks or attempts to conceal the recognition.

Ruben Gur, a neuropsychologist at the University of Pennsylvania, states that fMRI scans can reveal cognitive tasks when a subject recognizes a familiar situation, object, or person connected with any fraud, theft of intellectual property, and IT sabotage, no matter how hard he or she tries to conceal it [17]. This cognitive analysis technology could function as a hyper-accurate lie detector that is nearly impossible to deceive [35] [36]. For instance, an investigator could present a suspect with specific information such as of passwords, social security numbers, credit card information, other personal information, or other confidential corporate information [35]. This scientific technique allows intelligence operatives to focus their investigations on the suspects who actually commit crime and to determine if he or she has any un-authorized access to an organization's network or system before (See [22] [25] [26] [29] [33] [35]). On the other side, an information absent will provide support for the claims of innocence that employee is not guilty of committing any crime and has no knowledge specific to any data or any information [37]. This development will lead to speculations about the development of this neuro-imaging technology that could directly examine the malicious insider memories, intentions and its mind.

Interrogators will be able to confidently say that the fMRI told us this suspected employee lied about X or that he recognizes Y or fMRI picked him out as a malicious insider. This confidence that organizations will have in this neuroscience technique will be based on an aura of infallibility, scientific validity and objectivity [17]. Thus, unlike polygraph—which detects a person's emotional response to deception—fMRI measures person's decision to lie, as subjects cannot control their cerebral activity to avoid detection [38].

Thus fMRI can be used as a tool warranted in employee screening as not only has this neuro-imaging technology taken the attention of scientific communities and law enforcement agencies but it has also attracted interest of corporate world [38]. Two private firms: No Lie MRI and Cephos Corp trying to make the dream of perfect truth verification into a reality and have begun marketing since 2006. They offer high-tech lie detection services based on research comparing neuronal activation patterns [14].

The future of cloud provider may very well be under construction with this new approach that is becoming a reality for mining of knowledge from suspected employee to assess potential threats rapidly. However, there are still significant concerns must be addressed prior to moving this technology to real-world application [16]. In addition to the scientific challenges, advances in fMRI identify numerous, human rights, social, legal and general public concerns to the process of and the science behind it [39].

## VIII. EVIDENCE-BASED POLICIES AND GUIDELINES: A RELIABLE RESPONSE TO EMPLOYEES CONCERNS

In this research we are proposing best practices, recommendations and guiding principles to employ fMRI brain scanning technology during the questioning of suspects in cloud environment.

1. We recommend that, members involved in screening process must be made aware of the issues raised by this technology to develop best practices and efficient internal measures. These actions should address the development of policies and procedures relating to incidental findings within the doctrine of informed consent. Informed consent should be sought before scanning as the employees should be aware of the potential dangers. He/she should read, understand and sign an informed consent disclaimer to ensure that all the necessary requirements are met. This authority will give employee confidence and more control over the construction of their identities. Human experimentation without the consent of the subject is also a violation of human rights law [40]. To assure the protection, the fMRI scan process should undergo a complete government approval process to make reasonable assurance of employee's safety.

2. Training of investigators is one of the major challenges for the implementation of this tool. This training is necessary for the evaluation of screening centers to appropriately protect employees while allowing for scanning. Thus, only trained experts will be required to evaluate subjects and conduct the scan.

3. Experts are ethically obligated to report to the appropriate authorities when they have reason to believe that employee screening is coercive and violating human rights. They must ensure that if experts do not detect any abnormal behavior, the subject is not harmed. However, if an abnormality is detected, the results of the scan should be analyzed by other highly trained neuroscientists and possibly rectified [43].

4. It is also important that professionals involved in screening will be required to acquire security clearances. This shield will make it impossible for them to share the findings with colleagues in unclassified settings [43].

5. This research recommends that "Certificate of Confidentiality and Privacy" that can make a difference in the screening context. This certificate will allow the members who have access records to refuse to disclose identifying information at the civil, criminal, legislative, federal, state, or local level if the employee is not guilty. Disclosure of sensitive information could have adverse consequences on innocent person's reputation, employability as well as financial standing [41]. This document will particularly encourage employee to participate in the scanning process.

6. Any pre-employment screening process must be compatible with all relevant legislations, for instance,

employee's right must be protected by Article 8 [7] of the European Convention on the Protection of Human Rights and Article 12 [8] of the Universal Declaration of Human Rights. The process must implement the United Nations International Labor Organization (ILO) code of practice on the Protection of Workers' Personal Data (1996) [9] as well as European Union Guidelines 95/46 and 97/66 on data protection.

7. It is important that organizations must ensure the safety of the subjects through the systematic monitoring of the international law and human rights – including the United Nations Conventions against Torture, the International Covenant on Civil and Political Rights, and the Universal Declaration of Human Rights. The agencies must also consider the nuances of the Geneva Conventions as applied to suspected terrorists [42].

8. Question should be limited to a verification of the "real" or "personal" identity such as education, employment history, court records, credentials and other data associated with an employee. In screening process, the access to the results should be restricted for investigators in order to prevent the misuse of these preliminary data.

9. Uniformed personnel's and medical experts who are engaged in screening panel using fMRI must be held to account for their actions if they have violated human rights laws. Innocent employees or victim of this technology must be offered compensations, health care services and a formal apology to address ethical violations caused by this technology or by the professionals. A comprehensive federal investigation is required if the staff's trust in the ethical integrity of the security and medical profession being seriously compromised. If interrogators dismiss a subject for failing an fMRI scan test, they must be able to justify the action against him/her under the influence of a Human Rights Act, such as the European Convention on Human Rights (ECHR) or the UK Human Rights Act 1998. Furthermore, a policy can be introduced of only screening in case of suspicious activities. In this regard, drug testing examples of the Österreichischer Gewerkschaftsbund (ÖGB) in Austria, Deutscher Gewerkschaftsbund (DGB) in Germany, and the Confédération Générale de Travail (CGT) in France are suitable case studies for this approach [42].

10. Innovation in technology has been a key driver of change - the defense and security arenas are no exception. Similarly, members of elite unit should be well aware of current knowledge, novel literature, latest technologies, valuable processes and services about fMRI scanning for the purpose of developing image analysis to improve investigating methods. It is a major step forward in the action to our national interests that will continue to play a key role in the effectiveness of fMRI as a counterterrorism tool. We also recommend that government must push promising research on fMRI as they could meet our defense needs through collaboration with research sectors and universities to ensure a strong research base in this area. This action must be vibrant, inventive and innovative that looks most promising in screening neuroimaging. Investigators and neuroscientists must grasp the opportunities and adapt them quickly and effectively as this benefit is critical to our security and sovereignty [42].

---

[7] http://www.echr.coe.int/NR/rdonlyres/D5CC24A7-DC13-4318-B457-5C9014916D7A/0/ENGCONV.pdf

[8] http://www.un.org/en/documents/udhr/index.shtml#a12

[9] http://www.ilo.org/wcmsp5/groups/public/---ed_protect/---protrav/---safework/documents/normativeinstrument/wcms_107797.pdf

## IX. CONCLUSION

September 11th has marked an important turning point that exposed this new type of human threat that may pose a significant risk to critical infrastructure in cloud computing. This paper has explored different countermeasures that can be taken by an organization to protect themselves against this threat. The main objective for this work is to focus on advanced security strategies, frameworks, models, multi-layered security strategies and assessment methods linked to the overall architecture of this paper. Employee screening is central to this approach. The Insider Threat Study has also revealed a surprisingly high number of malicious insiders with prior criminal convictions when hired. Having access to complete source of employee history information is the only way the interest in performing due diligence to protect key assets and the nation can be served. At one hand it is beneficial for a general improvement which ultimately leads to higher productivity, better workers, increased efficiency and will provide an acceptable level of assurance for employees who have access to protectively marked critical assets and could alleviate the burden of mistrust. Furthermore, the aim of introducing this scanning technique is also deterrence from malicious activity of any kind. Indeed this approach may deter some high risk candidates with criminal/terrorist backgrounds from applying for the job at all, which moreover may even save money and time in the recruitment process. Similarly, there is a great expectation among scientists and counterterrorism agencies that workers at critical positions will also realize the urgency of the threat and the significance of neuroimaging application to national defense.

More significant, consideration must also be given to the government's purpose in subjecting the suspect to fMRI scan. It is important that the state's interest in interrogating a potential malicious insider in a high risk position must be justifiable, appropriate and will not curtail substantial civic liberty. Whether or not policy makers or civilized society can or should allow brain scanning is a matter that will continue to be debated for years to come. However given only the choice of deciding of whether or not scan an individual using fMRI technology when it may be possible to prevent mass causalities through this scan, the state ultimately has to make sensible decisions as necessary in order to save lives.